# The Improvement Program in Nonrelativistic Lattice QCD

Colin J. Morningstar [a] [*]

[a]Stanford Linear Accelerator Center, Stanford University, Stanford, California 94309

Progress in the improvement program in nonrelativistic lattice QCD is outlined. The leading radiative corrections to the heavy-quark mass renormalization, energy shift, and two important kinetic coupling coefficients are presented. The reliability of tadpole-improved perturbation theory in determining the energy shift and mass renormalization is demonstrated.

## 1. INTRODUCTION

The action in nonrelativistic lattice QCD [1,2] (NRQCD) must be engineered to reproduce the action of continuum QCD at low energies. The starting point in this process is a discretized version of the Schrödinger action. Interactions must then be added to systematically correct for relativity and finite lattice spacing errors. Lastly, the coupling strengths of these added interactions must be determined. One possibility is to treat these new couplings as adjustable parameters and tune them to fit certain experimental data; however, this tuning significantly reduces the predictive power of NRQCD simulations and is very costly and difficult. A better alternative is to compute the new couplings in terms of the fundamental QCD coupling $\alpha_s$ and the heavy-quark mass $M$ using perturbation theory. This is done by evaluating various scattering amplitudes in both QCD and lattice NRQCD and adjusting the couplings until these amplitudes agree at low energies. Since the role of these couplings is to compensate for neglected low-energy effects from highly-ultraviolet QCD processes, one expects that they may be computed to a good approximation using perturbation theory, provided that $M$ is large enough and the lattice spacing $a$ is small enough. This entire engineering procedure constitutes the improvement program in NRQCD.

In this talk, I report on progress being made in this improvement program. I present results for the leading radiative corrections to the mass renormalization, energy shift, and two important kinetic coupling coefficients. This work is an extension of previous calculations [3,4] and is being carried out in conjunction with ongoing simulations as part of the NRQCD collaboration [5-7].

## 2. LATTICE NRQCD

The lattice NRQCD action used here is written

$$S_Q^{(n)} = a^3 \sum_x \Biggl\{ u_0\, \psi^\dagger(x)\, U_4(x)\, \psi(x+a\hat{e}_4) \\ - \psi^\dagger(x)\left(1-\frac{aH_0}{2n}\right)^n (1-a\delta H)\left(1-\frac{aH_0}{2n}\right)^n \psi(x) \Biggr\},$$

where $n$ is a positive integer, $\psi(x)$ is the heavy-quark field, $U_\mu(x)$ is the link variable representing the gauge field along the link between sites $x$ and $x + a\hat{e}_\mu$, and $u_0$ is a mean-field parameter [8] defined in terms of the mean plaquette. The remaining operators are given by $H_0 = -\Delta^{(2)}/(2M)$ and $\delta H = \sum_{j=1}^{8} c_j\, V_j$, where the $V_j$ denote various interactions constructed from covariant, tadpole-improved difference operators and the cloverleaf chromoelectric and magnetic fields. Two of these operators are given by

$$V_1 = -\frac{(\Delta^{(2)})^2}{8M^3}\left(1+\frac{Ma}{2n}\right),$$
$$V_2 = \frac{a^2 \Delta^{(4)}}{24M},$$

where $a^{2m}\Delta^{(2m)} = \sum_{k=1}^{3}[u_0^{-1}[U_k+U_k^\dagger]-2]^m$. The parameter $n$ stabilizes the evolution of the quark

---

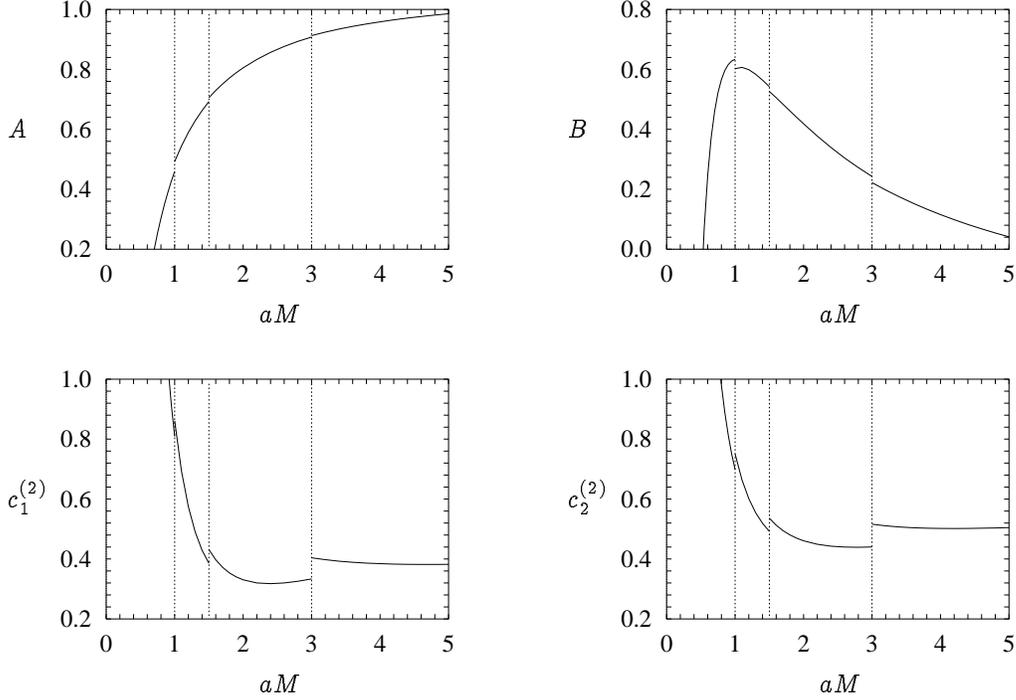

Figure 1. The energy shift parameter $A$, the heavy-quark mass renormalization parameter $B$, and the kinetic coupling coefficients $c_1^{(2)}$ and $c_2^{(2)}$, against the bare heavy-quark mass $aM$. For $aM > 3$, the stability parameter $n$ is set to unity; for $1.5 < aM < 3$, $n = 2$ is used; for $1 < aM < 1.5$, $n = 3$ is used; and for $0.5 < aM < 1$, $n = 6$ is used.

Green's function when $n > 3/(Ma)$, approximately. The coupling coefficients $c_j$ are functions of $\alpha_s$ and $aM$; at tree level, their values are unity.

## 3. HEAVY-QUARK SELF-ENERGY

The $O(\alpha_s)$ contribution to the heavy-quark self-energy has been determined. The small-$\mathbf{p}$ representation of the self-energy may be written:

$$a\Sigma(p) \approx \alpha_s \Big\{ f_0(w) + f_1(w) \frac{\mathbf{p}^2 a^2}{2Ma} + f_2(w) \frac{(\mathbf{p}^2)^2 a^4}{8M^2 a^2} + f_3(w) \, (\sum_{j=1}^{3} p_j^4 a^4) + \ldots \Big\},$$

where $w = -ip_4 a$ and $f_m(w) = \sum_{l=0}^{\infty} \Omega_m^{(l)} w^l$. The on-mass-shell quark satisfies a dispersion relation

$$\omega_0(\mathbf{p}) \approx \frac{\mathbf{p}^2}{2M_r} - \frac{(\mathbf{p}^2)^2}{8M_r^3} - \alpha_s \delta\omega(\mathbf{p}) + \ldots,$$

where $M_r = Z_m M$ is the renormalized mass with $Z_m = 1 + \alpha_s B$, $B = \Omega_0^{(0)} + \Omega_0^{(1)} + \Omega_1^{(0)}$, and

$$a\delta\omega(\mathbf{p}) = \Omega_0^{(0)} + \Omega_3^{(0)} \, (\sum_{j=1}^{3} p_j^4 a^4) + \Omega_\omega \frac{(\mathbf{p}^2)^2 a^4}{8M^2 a^2},$$

$$\Omega_\omega = \Omega_0^{(0)} \left(1 + \frac{2}{Ma}\right) + 2\Omega_0^{(1)} \left(1 + \frac{1}{Ma}\right)$$
$$+ 2\Omega_0^{(2)} + \Omega_1^{(0)} \left(2 + \frac{3}{Ma}\right) + 2\Omega_1^{(1)} + \Omega_2^{(0)}.$$

The sole effect of the $O(\alpha_s)$ corrections in QCD is to renormalize the quark field and mass. If lattice NRQCD is to reproduce the low-energy physical predictions of full QCD, then the $O(\alpha_s)$ corrections to the heavy-quark propagator in lattice NRQCD must also do no more than renormalize the heavy-quark field and mass at low momenta. This will be true if $\delta\omega(\mathbf{p}) = 0$; that is, if $\Omega_0^{(0)} = \Omega_3^{(0)} = \Omega_\omega = 0$. Alternatively, since the constant term $\Omega_0^{(0)}$ represents an overall energy shift, one could require only $\Omega_3^{(0)} = \Omega_\omega = 0$ and then simply shift the energies obtained in simulations by an amount $-\alpha_s A/a$ for each heavy quark, where $A = -\Omega_0^{(0)}$. Setting $\Omega_3^{(0)} = \Omega_\omega = 0$ then determines the kinetic coupling coefficients $c_1^{(2)}$ and $c_2^{(2)}$, where $c_j = 1 + \alpha_s c_j^{(2)}$.

Table 1
Comparison of the $\Upsilon$ mass $M_\Upsilon^{(a)}$ expected from perturbation theory using Eq. 2 and the mass $M_\Upsilon^{(b)}$ obtained by fitting simulation results to the dispersion relation in Eq. 1. Values for $\alpha_V(q^*)$ are obtained by fitting simulation measurements of the mean plaquette to its perturbative expansion [5].

| $aM$ | $aE_{NR}(\Upsilon)$ | $Z_m - 1$ | $aE_0$ | $aM_\Upsilon^{(a)}$ | $aM_\Upsilon^{(b)}$ |
|------|---------------------|-----------|--------|---------------------|---------------------|
| 2.00 | 0.444(1) | $.43\alpha_V\left(.67a^{-1}\right) = .14$ | $.81\alpha_V\left(.81a^{-1}\right) = .22$ | 4.53 | 4.45(5) |
| 1.80 | 0.451(1) | $.46\alpha_V\left(.67a^{-1}\right) = .15$ | $.77\alpha_V\left(.81a^{-1}\right) = .21$ | 4.15 | 4.10(5) |
| 1.71 | 0.455(1) | $.48\alpha_V\left(.67a^{-1}\right) = .15$ | $.75\alpha_V\left(.81a^{-1}\right) = .21$ | 3.98 | 3.95(5) |

## 4. RESULTS

The values for $A$, $B$, $c_1^{(2)}$, and $c_2^{(2)}$ after tadpole improvement are shown in Fig. 1. The values for these parameters before tadpole improvement are typically very large. In the tadpole improvement, the perturbative expansion for the mean field parameter $u_0 = 1 + \alpha_s u_0^{(2)}$, where $u_0^{(2)} = -1.047$, is applied. The renormalized coupling $\alpha_V(q^*)$ of Lepage and Mackenzie [8] is used to deduce values for $\alpha_s$; fitting simulation measurements of the mean plaquette to its perturbative expansion specifies $\alpha_V(q^*)$ [5]. The computation of the scales $q^*$ for the above parameters is nearly completed.

It is possible to check the reliability of the perturbative determination of the energy shift and mass renormalization. Recent NRQCD simulations [5-7] have produced values for the rest energy $aE_{\mathrm{NR}}(\Upsilon)$ of the $\Upsilon$. In addition, energies at small but non-zero momentum values have been obtained and fit to the dispersion relation

$$E_\Upsilon(\mathbf{p}) = E_{\mathrm{NR}}(\Upsilon) + \frac{\mathbf{p}^2}{2M_\Upsilon^{(b)}} + \cdots, \quad (1)$$

yielding a nonperturbative estimate $M_\Upsilon^{(b)}$ of the $\Upsilon$ mass. This mass can be compared with that expected from perturbation theory, namely,

$$aM_\Upsilon^{(a)} = 2(aM_r - aE_0) + aE_{\mathrm{NR}}(\Upsilon), \quad (2)$$

where $aE_0 = \alpha_V(q^*)A$. The results of this comparison are shown in Table 1. The masses agree down to the $1-2\%$ level.

## 5. CONCLUSION

The leading radiative corrections to the heavy-quark mass renormalization, energy shift, and two important kinetic coupling coefficients were presented. The results underscore the reliability of tadpole-improved perturbation theory in the improvement of nonrelativistic lattice QCD.

## 6. ACKNOWLEDGEMENTS

This work was supported by the NSERC of Canada, the U.S. DOE, Contract No. DE-AC03-76SF00515, and the UK SERC, grant GR/J 21347.